\documentclass{aa} 
\usepackage{graphicx}
\usepackage{txfonts}
\begin{document}
\title{The pulsed X-ray light curves of the isolated neutron star 
RBS1223\thanks{Based on observations obtained with XMM-Newton, an ESA
  science mission with instruments and contributions directly funded by ESA
  Member States and NASA}} 

\author{A.D. Schwope\inst{1}
          \and
        V. Hambaryan\inst{1}
          \and
        F. Haberl\inst{2}
          \and
        C. Motch\inst{3}}

\offprints{A. Schwope}

\institute{Astrophysikalisches Institut Potsdam, An der Sternwarte 16,
           14482 Potsdam, Germany
    \and
             Max-Planck-Institute f\"ur Extraterrestrische Physik,
              Giessenbachstr., 85748 Garching, Germany
\and
Observatoire Astronomique, CNRS UMR 7550, 11 rue de l'Universit\'e, 67000
              Strasbourg, France
             }

\date{Received ; accepted May 26, 2005}

\abstract{We present multi-epoch spectral and timing analysis of the
isolated neutron star RBS1223. 
New XMM-Newton data obtained in January 2004 confirm the spin
period to be twice as long as previously thought, $P_{\rm spin} =
10.31$\,s. The combined ROSAT, Chandra, XMM-Newton data (6 epochs)
give, contrary to earlier findings, no clear indication 
of a spin evolution of the neutron star.
The X-ray light curves are double-humped with pronounced hardness ratio
variations suggesting an inhomogeneous surface temperature with two spots 
separated by about $\sim$160\degr. 
The sharpness of the two humps suggests a mildly relativistic star with a
ratio between $R_{\rm ns}$, the neutron star 
radius at source, and  $r_{\rm S}$, the Schwarzschild-radius, 
of $R_{\rm ns}/r_{\rm S} > 3.5$. 
Assuming Planckian energy distributions as local radiation sources,
light curves were synthesized which were found to be in overall
  qualitative agreement with observed light curves in two different 
energy bands. The temperature distribution used was based on the 
crustal field models by Geppert et al.~(2004) 
for a central temperature of $T_c = 10^8$\,K and an
external dipolar field of $B \sim 10^{13}$\,G. This gives a mean atmospheric
temperature of 55\,eV. A much simpler model with two homogeneous spots with
$T_\infty = 92$\,eV, and 84\,eV, and a cold rest star, $T_{\rm star,\infty} <
45$\,eV, invisible at X-ray wavelengths, was found to be similarly
successful.
The new temperature determination and the new
$\dot{P}_{\rm spin}$ suggest that the star is older than previously
thought, $T \simeq 10^{5\dots6}$\,yrs. The model-dependent distance 
to RBS1223 is estimated between 76\,pc and 380\,pc 
(for $R_{\rm ns} = 12$\,km).
}
\authorrunning{A.D. Schwope et al.}
\titlerunning{X-ray light curves of RBS1223}
\maketitle
%

\section{Introduction}
RBS1223 belongs to the small elusive group of X-ray dim isolated neutron stars 
(XDINs) discovered in the ROSAT all-sky survey (RASS). The currently known seven
systems share the properties: 
soft, blackbody-like X-ray spectrum, no radio emission, no
association with a supernova remnant (for a review see Haberl 2004). 
Five objects are X-ray pulsars with spin periods in the range 3.45 to 11.37
s. The spectra of all those stars were fitted originally with pure Planckian
spectra. While the spectrum of the brightest and best-studied, RX\,J1856.4--3754, 
is still compatible with a pure black-body model (Burwitz et al.~2003),
recent XMM-Newton observations of RXJ0420.0-5022, RXJ0720.4-3125,
RXJ1605.3+3249 (= RBS1556) and RBS1223 (= 1RXS J13048.6+212708), however,
reveal significant deviations from the Planckian shape of the X-ray spectrum. 
These were tentatively identified with proton cyclotron absorption lines in
fields of a few times $10^{13}$\,G (Haberl et al.~2003, 2004, van. Kerkwijk et
al.~2004). 
The nature of these XDINs is not clear. Their measured spin periods
and moderate spin-down rates are suggestive of middle-aged neutron stars on
their cooling tracks, their high surface temperatures seem to favour younger
ages. The origin of their X-ray emission and their atmospheric composition
is still under debate.

RBS1223 was discovered in the course of an optical identification program of
the more than 2000 RASS X-ray sources at high galactic latitude with count
rate CR $> 0.2$\,s$^{-1}$ (Schwope et al.~2000). The RASS error circle of
source \#1223 in this catalogue was
without obvious optical counterpart at the DSS limit. Subsequent ROSAT HRI 
observations gave an  improved X-ray position. Deep Keck imaging, $R_{\rm lim} 
\sim 26^m$, remained without optical counterpart, suggesting a high ratio 
$f_{\rm X}/f_{\rm opt} > 10^4$, 
which excluded anything else than a neutron star
as the X-ray source (Schwope et al.~1999, paper 1). 
Follow-up Chandra observations revealed a periodically modulated X-ray signal
with an  oscillation period of about 5.16\,s (Hambaryan et al.~2002, paper 2).
Retrospectively, we also found those oscillations in the HRI data. The derived 
spin-down rate of the neutron star of $\dot{P}\sim 1\times
10^{-11}$\,s\,s$^{-1}$ implied an ultra-high magnetic field, $B_{\rm dip} >
10^{14}$\,G, and a characteristic age of around $10^4$\,years, if interpreted
as due to magnetic dipole braking. The decay of the field would also
provide an explanation for the X-ray luminosity of the source.
However, at the implied young age the source would
not have been able to travel far away from its birth place. The
non-detection of a close-by supernova remnant at radio wavelength or in X-rays
is then at least puzzling. 

We have obtained XMM-Newton observations at a first epoch in 2001
(satellite revolution rev 377)
with all three EPIC cameras and both RGS'. A timing and spectral analysis of these
data together with a second epoch observation 
was published by Haberl et
al.~(2003, paper 3). The much better photon statistics showed a double-humped
X-ray light curve and a spin period of twice the previously reported 
value, $P_{\rm
spin} = 10.31$\,s. Simple blackbody models did not fit the mean X-ray spectra
at both epochs. A much better representation of the observational data was achieved
after including a Gaussian-shaped absorption line superposed on a black-body spectrum. 
Under the assumption that the observed line is caused by a proton cyclotron
line, the observed position indicates a field strength 
of $2 - 6 \times 10^{13}$\,G. This value would still imply a detectable spin-down
rate although much smaller than derived in paper 2. If the line would be an 
electron cyclotron line, the implied field strength would be of order $\sim
10^{11}$\,G. One would not expect to be able to detect any spin variation
caused by the field decay over the comparatively short period of time 
that this object is observed now (1996 -- 2004). 

In this paper we present new data from calibration observations of RBS1223
obtained with XMM-Newton in 
January 2004 and GO time data obtained with the Chandra LETG 
on March 30, 2004.  
We combine all available data sets in order to update the spin history of the
object. We describe a model which mimics
the pulsed X-ray light curves of RBS1223. It
allows to determine the sizes and temperatures of the spots on the neutron
stars surface and gives also a distance estimate. 

\section{Observations and analysis}
XMM-Newton observed RBS1223 meanwhile on three occasions. At the first epoch
on Dec.~31, 2001 (XMM-Newton revolution (= rev) 377), RBS1223 was observed as
a GO-target, for the 
following two epochs on Jan 1., 2003 (rev 561) and Dec.~30, 2003 (rev 743), it
was observed as a calibration target by the XMM-Newton SOC. The first two
observations were described already in paper 3, the observation in rev.~743 is
almost a carbon copy of that in rev.~561, although with somewhat longer
exposure time (32\,ksec instead of 28\, ksec). 
For details of the data reduction see paper 3. For the
purpose of the present paper we have re-done the basic data reduction steps
with the most recent version of the SAS (v6.0.0) commonly applied to all
XMM-Newton data sets. 
A Chandra LETG observation was performed on March 30, 2004 for a total
  exposure of 91 ks. 
For the purpose of the present paper we make use of the photon arrival times
of 82 ks low-background time. The spectral distribution of the photons will
be investigated and published separately (Haberl et al.~, in preparation).

\begin{table}[ht]
\caption{Spin period of RBS1223 at given epoch with given instrument. All
  periods listed are newly derived values. 
\label{t:spin}} 
\begin{tabular}{lll}
  Epoch MJD & Period & Instrument\\
(days)  &       (sec)  &      \\
\hline
50824.21390  &   $ 10.31227_{-0.00011}^{+0.00012}$ &  Rosat HRI\\
51719.51835  &   $ 10.31307_{-0.00048}^{+0.00060}$ &  Chandra ACIS-S\\
52274.25598  &   $ 10.31253_{-0.00029}^{+0.00030}$ &  XMM PN AO1\\
52640.43264  &   $ 10.31250_{-0.00020}^{+0.00020}$ &  XMM PN CAL1\\
53003.46683  &   $ 10.31258_{-0.00018}^{+0.00018}$ &  XMM PN CAL2\\
53095.33034  &   $ 10.31250_{-0.00012}^{+0.00004}$ &  Chandra LETGS\\
\hline
\end{tabular}
\end{table}
\begin{figure}[th]
\resizebox{\hsize}{!}{\includegraphics{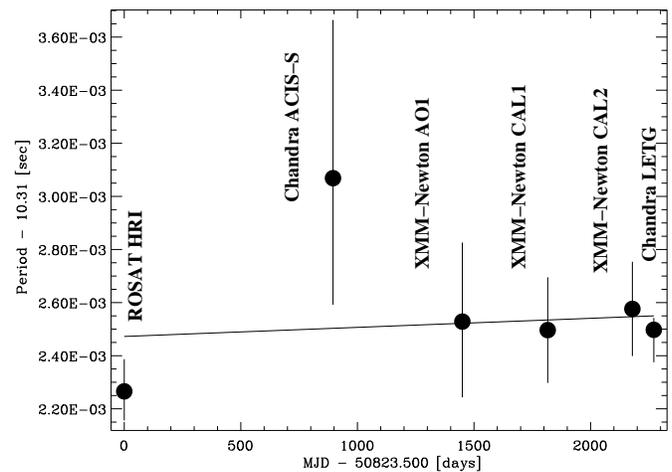}}
\caption{The period history of RBS1223 based on X-ray observations at
six epochs. The straight line represents a weighted linear fit to all data. 
\label{f:perhist}
}
\end{figure}

\section{Revised period and updated spin history of RBS1223}
\begin{figure}[t]
\resizebox{0.95\hsize}{!}{\includegraphics[clip=]{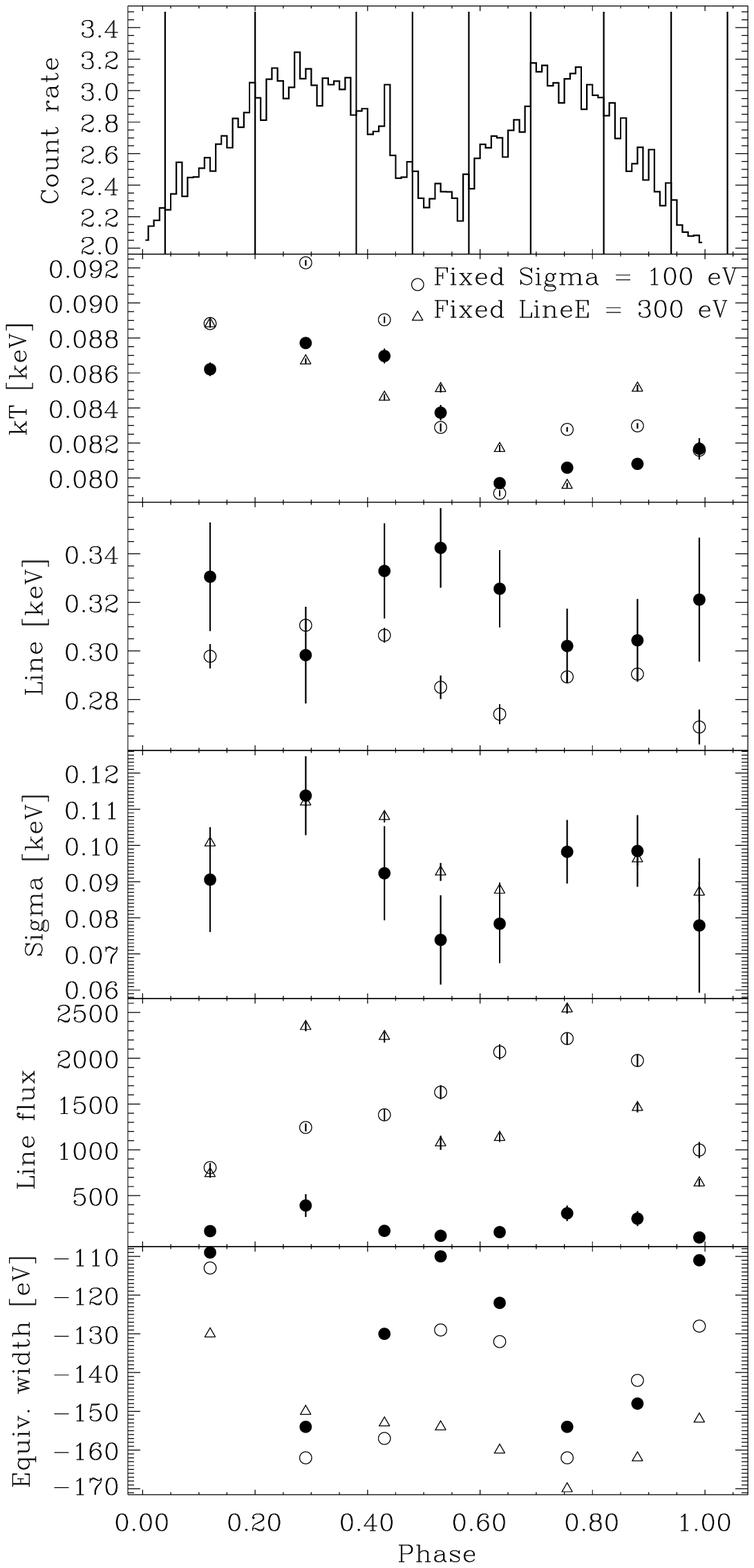}}
\caption{Phase-averaged light curve and phase-resolved spectroscopy of the
XMM-Newton observations of RBS1223 performed in revs.~561 and 743,
respectively. The light curve in the top panel  
encompasses the spectral band 0.12 -- 1.2\,keV. Vertical lines indicate the
boundaries of data groups for subsequent spectral analysis. In the panels
below are shown the best-fit blackbody temperature, the centre, the width, the
integrated flux and the equivalent widths of the Gaussian absorption line,
respectively. 
All these parameters were determined by unconstrained
fits (indicated by filled circles), by fits with fixed Gaussian line position
(open triangles), and by fits with fixed 
Gaussian line width (open circles). 
\label{f:lc}
}
\end{figure}

In order to determine most likely spin periods at the different epochs
we re-analysed all available data sets of RBS1223 obtained with ROSAT,
Chandra and XMM-Newton using the most recent calibration files and software
updates. We concentrated the period search on an interval around the
value identified as the most likely spin period in Paper 3 at 10.31\,s.
The algorithm used was described already in Paper 2. 
In short, to evaluate the most probable frequency and its 
uncertainty in a rigorous way, we analyzed these data with 
the method of Gregory \& Loredo (1992) based on the Bayesian 
formalism. This approach is free of any assumption about the pulse shape and 
results in an accurate parameter estimation
using the probability distribution function of frequencies.
The most recent data obtained with XMM-Newton confirm the 10.31 periodicity as
the true spin period. This identification rests on the phase-folded light and
hardness ratio curves which display a double-humped structure with unequal
count rate and spectral hardness (see Fig.~\ref{f:lc}). 

We re-analysed the ROSAT and the Chandra AO1 observations and found evidence
for periodic variability in the interval between 10.3110\,s and 10.3116\,s
also in those datasets, although with lower significance than at the 5.16\,s
periodicity. The probability that we have a periodic signal in the mentioned
time interval based on the ROSAT data is 23\%. For all other cases the
probability is almost 100\%. The measured spin
periods as derived for the six epochs between June 1997 and March 2004 are
listed in Tab.~\ref{t:spin} and displayed in Fig.~\ref{f:perhist}. The values 
given there are the most likely periods (highest peaks in the odds' ratio
periodogram), 
while the errors given are $1\sigma$ uncertainties (68\% confidence region, 
"posterior bubble"; see Gregory \& Loredo 1992 and Paper 2). 
Weighted linear fits to all data in Tab.~\ref{t:spin} were
performed using different weighting schemes. We used either pure statistical
errors taking into account the larger of the two values listed in the table,
or gave weights to individual data points according to the time resolution
element of the observation, the number of detected photons or the probability
of detection of the periodic signal via the odds ratio.
Using statistical weights only, $\dot{P} = (8.8 \pm  0.4) \times
10^{-13}$\,s\,s$^{-1}$ is derived. If weights according to the observed
number of photons and the significance of detection are given in addition, the
resulting $\dot{P} = (4.0 \pm  0.4 ) \times 10^{-13}$\,s\,s$^{-1}$ (fit shown
in Fig.~\ref{f:perhist}). 
Both fits indicate a small spin-down of the star.
The evidence for a true spin-down, however, is rather weak. 
It rests mainly on the ROSAT HRI 
observation in June 1997, were only 521 photons were collected. 
When the first data point from
Tab.~\ref{t:spin} is omitted, a weigthed fit to the remaining data indicates
even a spin-up of the neutron star, 
$\dot{P} = (-3.4 \pm 0.4) \times 10^{-13}$\,s\,s$^{-1}$. 
We conclude, that the presently available data are insufficient to determine
the spin history unequivocally. This is due to the fact that the period 
determination for each observation has insufficient accuracy to connect
subsequent observations without cycle count alias. However, our original
value for the spin-down of RBS1223 is clearly ruled out. 
 
\section{Phase-resolved X-ray spectroscopy}
With the sufficiently large number of photons collected by XMM-Newton 
spinphase-resolved low-resolution X-ray spectroscopy is possible after
epoch-folding and phase-averaging of the original data. Here and in the
following section we use the EPIC-pn data of the two calibration observations
(CALs) of RBS1223 only (revs.~561 and 743) in order to eliminate remaining
calibration uncertainties between the redistribution matrices of the two CAL
and the GO observations. The two CAL observations were performed in
normal full frame mode with a frametime of 73 ms.

The background-corrected mean countrate detected in the pn was 
$2.896\pm0.012$\,s$^{-1}$ and  $2.929\pm0.012$\,s$^{-1}$ 
for the CALs, respectively. We could not detect any
long-term photometric variation among the observations with XMM-Newton. 
In paper 3 we have already shown that the Chandra data from AO1
are compatible with the XMM-Newton AO1 data, i.e.~long-term changes of the
X-ray brightness between 2000 and 2004 are insignificant.

Using the 72151 and 69435 photons detected in the two CALs we created
phase-folded 
X-ray light curves for the two epochs. We then determined a relative
phase-shift by correlating the two light curves and finally created one common
phase-averaged light curve for both CALs with 100 phase bins in the energy
range 0.12 -- 1.2\, keV. The result is
shown in Fig.~\ref{f:lc}, top panel.

For our spectral analysis we defined eight phase intervals indicated by
vertical lines in Fig.~\ref{f:lc}: the centres of the
main and secondary pulses; the centres of the main and secondary
minima; the rise and 
the fall to the main and secondary pulses, respectively. 
We adopted the same spectral model as in Paper 3,
i.e.~we used a blackbody plus Gaussian absorption line absorbed by the same
amount of interstellar matter 
(represented by the column density $N_{\rm  H}$). The spectral analysis was
performed within XSPEC (version 11.3).
Three different flavours of our adopted model were tried, the first
with unconstrained parameters, a second with the centre of the
Gaussian absorption line kept fix and a third with the width of the
line kept fix. 

\begin{figure}[t]
\resizebox{\hsize}{!}{
\includegraphics[clip=]{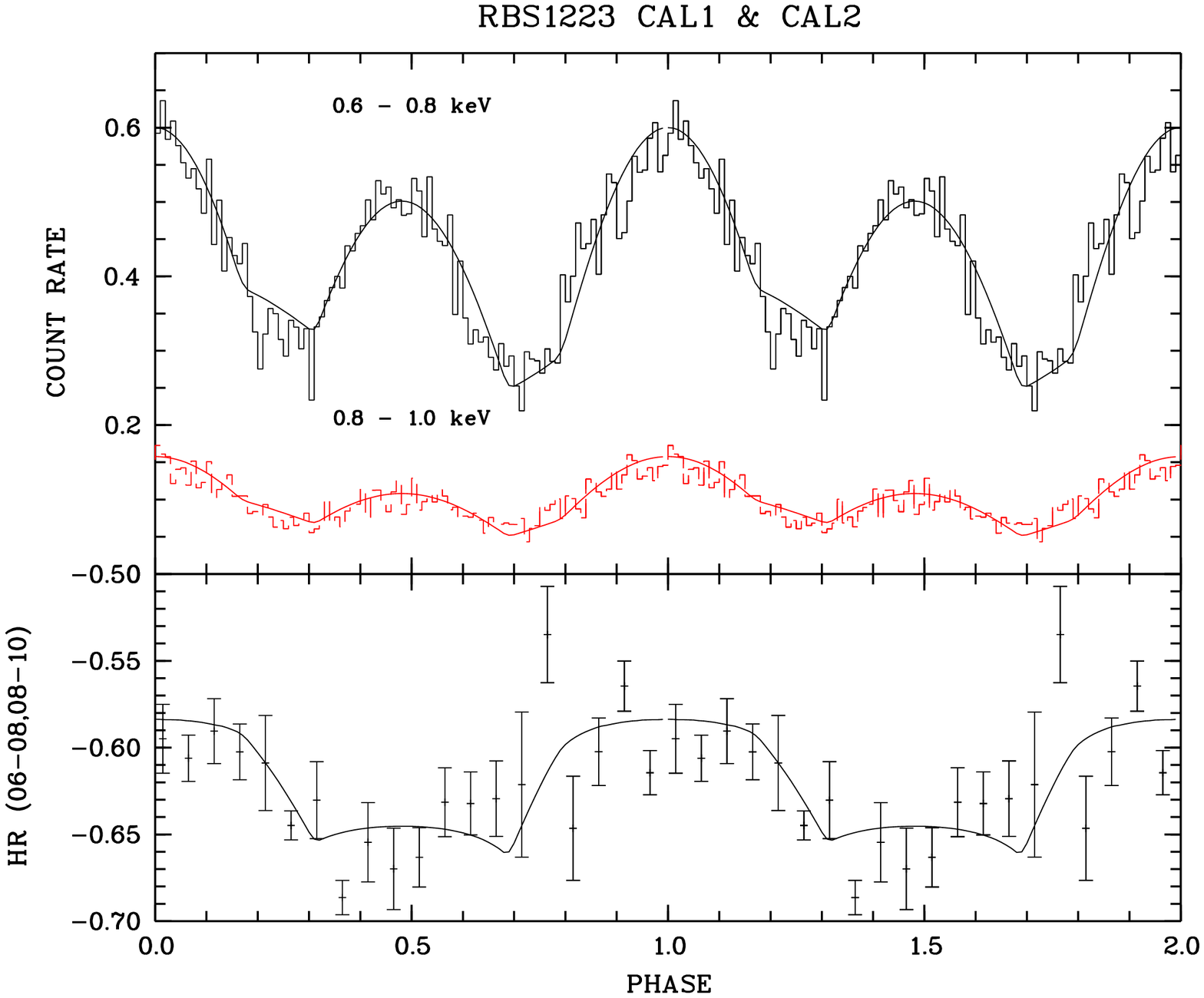}}
\caption{Soft X-ray light curves of RBS1223 in
energy bands 0.6 -- 0.8\,keV, and 0.8 -- 1.0\,keV (upper panel) 
and corresponding hardness ratio (lower panel). The solid lines are
predicted light curves based on the two spot model as described in the
text.
\label{f:2sm}
}
\end{figure}

Visual inspection of the $(O-C)$ residuals as well as an $F$-test revealed,
that the observed spectrum at any given phase cannot be fitted with just one
spectral (black-body) component. The inclusion of the Gaussian absorption line
resulted in a satisfactory fit to all 8 spectra. 
The temperature drops from about 90\,eV in
the first hump to about 82\,eV in the second hump, somewhat dependent on the
chosen model. The spectral parameters are depicted in the bottom panels of
Fig.~\ref{f:lc}. We found indication for a small change of the line
width, being broadest in the light curve humps.
Similarly, the absolute flux and the equivalent width of the Gaussian
absorption line is likely to be highest in the humps. Our search for 
a variation of the line centre remained inconclusive.

\section{Light curve analysis}
The X-ray light curves of RBS1223 have a markedly double-humped
shape with pronounced hardness ratio, i.e. spectral variations. 
The maxima and minima have unequal height. The maxima are separated by
0.47, the minima by 0.43 phase units, respectively. 
The occurrence of such X-ray oscillations is related to the presence
of a magnetic field. Shibanov et al.~(1992), Zavlin et al.~(1995) and Zavlin
\& Pavlov (2002) have shown that the simple presence of a sufficiently strong
magnetic field causes anisotropic radiation. The observed pronounced spectral
variations clearly show a temperature inhomogeneity. 

\begin{figure}[t]
\resizebox{\hsize}{!}{
\includegraphics[clip=]{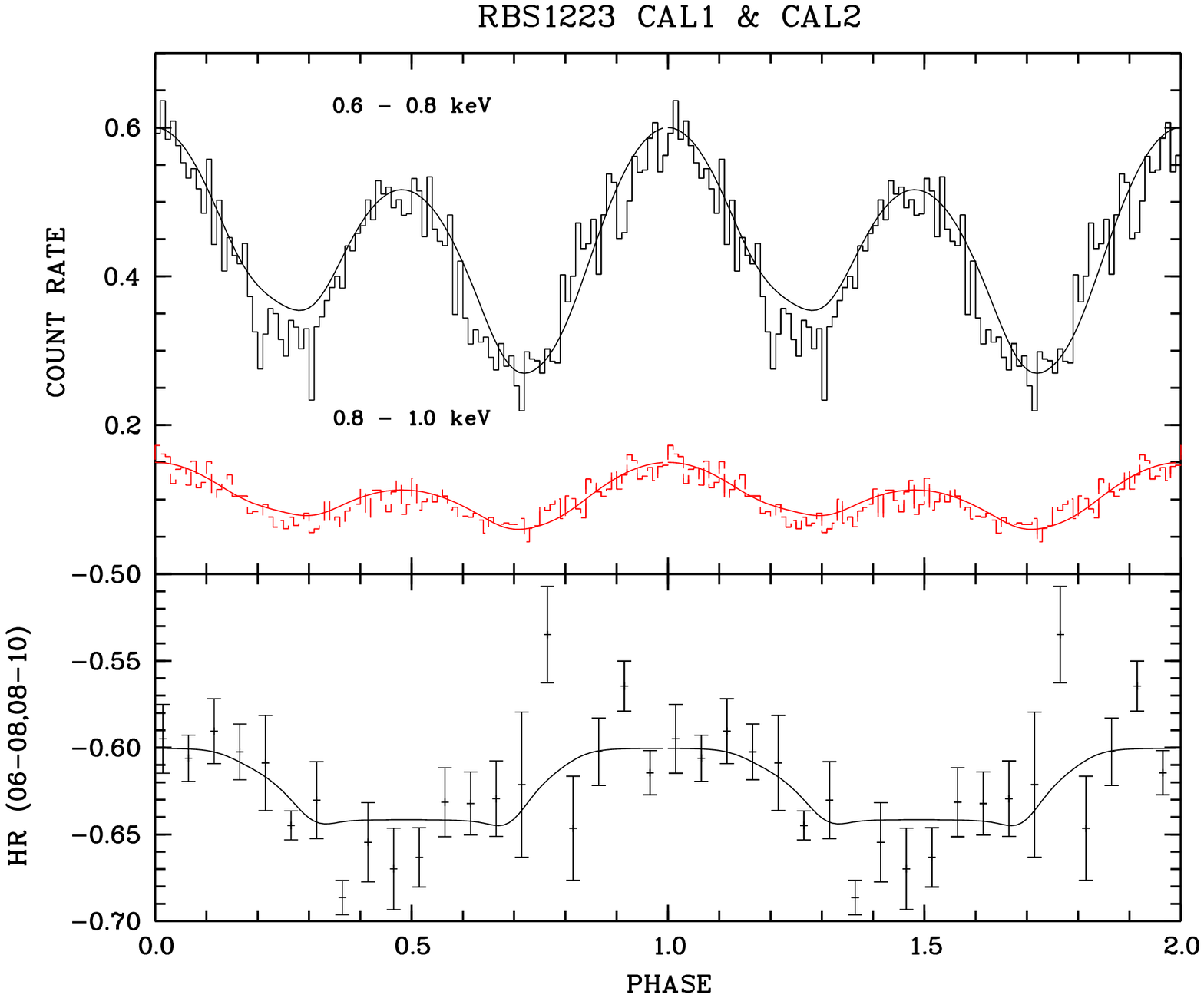}}
\caption{Observed data as in Fig.~\ref{f:2sm}, the model light and
  hardness ratio curves are based on crustal field temperature
  profiles according
  to Geppert et al.~(2004). Model parameters are described in the text. 
\label{f:cfm}}
\end{figure}

We regard temperature inhomogeneities as the main source of the observed
variations. Here we make an attempt to model the light curves assuming an
inhomogenous temperature distribution over the surface of the neutron
star. This is a first-order approximation, a full model would have taken into
account anisotropic radiation according to the local magnetic field in a
surface element. Those models are not yet available for arbitrary
  angles between the surface normal and the orientation of the
  field. As a further complication the numbers of degrees of freedom
  of the models would be raised by an order of magnitude. Without a
  proper normalization scheme and perhaps even much better data in terms of
  spectral and time resolution it would be very difficult to discern
  different models. For the time being we think that the simple approach we
  are following in this paper is justified by the data. 

The surface of the star
is tiled into small areas using constant steps in latitude and
longitude. A temperature is assigned to each of the surface
elements. In model 'A' two spots of circular shape were located 
at certain positions, in model 'B' a temperature distribution
according to the recent crustal field models by Geppert et al.~(2004)
is assumed. We assume black-body radiation from each of the elements
for the assumed temperature. The modeled photon spectrum is folded
through the response of the XMM-Newton EPIC-pn camera. The predicted
count-rate at any given time (phase) is the summed rate
of all visible tiles at that phase, with an appropriate
fore-shortening factor for each of the tiles. A similar approach
was used by Page (1995) who gives a full description of the details.

The number of visible tiles at any given time depends on the
inclination $i$ of the observer (angle between the rotation axis and
the line of sight at infinity), the angle between the magnetic axis
and the rotation axis $\vartheta$, and the compactness of the neutron
star, parameterized by the ratio $r_{\rm gr} = R_{\rm ns} / r_{\rm S}$
($R_{\rm ns}$: radius of the neutron star (rest frame), 
$r_{\rm S}$: Schwarzschild-radius). 
The radius and temperature in the observers frame at infinity are related
to the quantities in the star's rest frame by $R_\infty = R_{\rm ns} (1+z)$,
$T_\infty = T_{\rm eff} / (1+z)$, $(1 + z)^{-1} = \sqrt{ 1 - 1/r_{\rm
    gr}}$ ($z$: gravitational redshift). 

The relevant parameters for modelling the light curves are then the
radius and temperature distribution of the star, the compactness, the
distance, the geometry, and the column density $N_{\rm H}$ of the interstellar
absorption. Due to the unknown physical nature of the absorption
feature at 0.3 keV, we restricted our light-curve analysis to the spectral
regime above 0.6\, keV, which is mainly unaffected by this
feature. Throughout the analysis we fixed $N_{\rm H}$ at the value
derived in the previous section, $N_{\rm H} = 4.8 \times
10^{20}$\,cm$^{-2}$. We perform our analysis in two energy bands, $B_1
= 0.6-0.8$\,keV, $B_2 = 0.8-1.0$\,keV; the corresponding hardness ratio
is defined in the usual manner as $HR = (CR_2 - CR_1)/(CR_1+CR_2)$,
$CR_{1,2}$ being the count rates in the two defined energy bands.
For any given $N_{\rm H}$ the hardness ratio is a direct measure of
the blackbody temperature.

Beloborodov (2002) has studied the bending of light in the vicinity of
compact objects. He derived a simple approximate relation between the
local angle of photon emission, $\alpha$, and the escape direction,
$\psi$: $1 -\cos\alpha = (1-\cos\psi)(1-r_{\rm gr})$. It has a
maximum deviation of 3\% for $r_{\rm gr} = 3$ with respect to the
exact theory (Pechenick et al.~1983). The visible fraction of the star
is derived from this relation by setting $\alpha = 90\degr$ (see his
Fig.~1). We use this approximation in
order to determine the visible fraction of the neutron star's surface
as well as the foreshortening angle of individual surface
elements. 

Beloborodov also studied the possible types of light curves as a
function of $i$ and $\vartheta$ if two spots are located diametrically
opposite on the magnetic axis. His class III is describing the current
situation of RBS1223 with the two spots rotating subsequently into view.
This type of light curve is observed, when $\cos(i+\vartheta) < -
r_{\rm S}/(R-r_{\rm S})$, i.e.~when both angles $i$ and $\vartheta$
are sufficiently large. Although in principle a large 
range of combinations between $i$ and $\vartheta$ 
produces light curves of type III, the pulsed fraction is maximized
for $i = \vartheta$, which we assume in the following.

We ran a number of simulations exploring the parameter space searching
for an optimum solution by a fit-per-eye. Our finally accepted 
light-curve solution giving best overall qualitative agreement with
the observed data is shown in Fig.~\ref{f:2sm}.
The parameters in our model 'A' are constrained by the following
observed features. 

The pulsed fraction of the light curve is rather
high, $p = (CR_{\rm max}-CR_{\rm min})/CR_{\rm max} \sim 52\%$ in the 
soft band $B_1$, which constrains the compactness of the star. For $r_{\rm
gr} < 3.5$ a too large fraction of the star becomes visible at any given
time. This damps the amplitude of the variation below the observed value. 
In the following we assume $r_{\rm gr} = 4$. This retrospectively justifies
the use of Beloborodov's approximative formula for the description of light
bending, which is applicable only to not too compact objects. 

The observed large pulsed fraction also requires a
high inclination $i$ (and a high $\vartheta$ for our assumed geometry with 
$i=\vartheta$), $i > 75\degr$, although one can trade to some extent a higher
compactness
parameter against a lower inclination and vice versa. The model light
curve shown in Fig.~\ref{f:2sm} was computed for $i=\vartheta=80\degr$
and $r_{\rm gr} = 4$.

The temperature of the main spot is fixed by the observed hardness
ratio in the main hump of the light curve to $T_{1,\infty} = 92$\,eV. 
The width of the
main hump constrains the maximum extent of the spot, the full opening
angle of the main spot is of order $2 \theta_1 \sim 8\degr$. 

The location of the second spot is constrained by the observed separation
between the maxima and minima of the light curve. We locate the second
spot at an offset angle of $\kappa = 20\degr$ with respect to the magnetic
axis and at an azimuth of about $20\degr$. The offset $\kappa$ is not very
well constrained by the observations. It could become  larger
if the spot is located closer to the meridian through the rotation and
the magnetic axis. Under extreme circumstances, $\kappa$ can become as large
as $50\degr$, although the fit then clearly deteriorates, 
hence we estimate the minimum separation between the two spots to be
$\sim$130\degr. 

The temperature of the second spot is determined by the observed spectral
hardness in the pulse maximum, $T_{2,\infty} = 84$\,eV. The width of the second
pulse and the relative brightness of the second with respect to the
first spot then suggest a full opening angle somewhat larger as for
the main pulse, $2\theta_2 \sim 10\degr$.

The contribution of the remaining surface in this model is negligible,
the light and hardness ratio curves are almost completely
determined by emission from the two spots. This contrains the
atmospheric temperature to $T_{\rm star,\infty} < 42$\,eV. The maximum
possible $T_{\rm star,\infty}$ is 50\,eV. This limit is derived under
the extreme assumption $r_{\rm gr} \gg 1$. In this non-relativistic
limit the two spots rotate alternating into view and the emission
between the two humps is entirely due to photospheric
emission.
 
\begin{figure}[t]
\resizebox{\hsize}{!}{
\includegraphics[clip=]{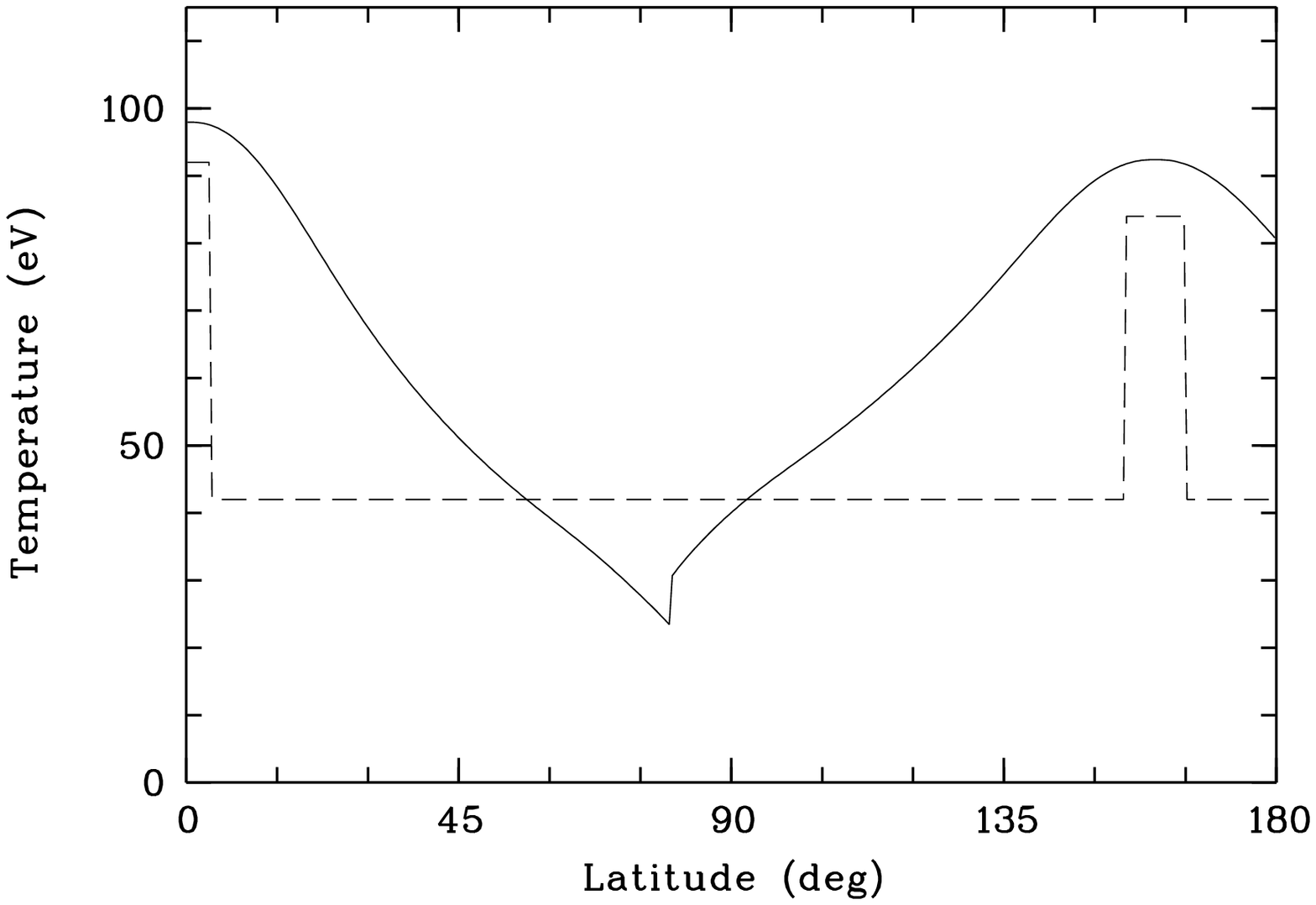}}
\caption{Temperature distribution as a function of magnetic colatitude for
  models 'A' (dashed line) and 'B' (solid line), respectively. The
  temperatures are given in eV. The stellar temperature of model 'A' is an
  upper limit as derived from the light curve modeling. This limit violates
  the constraint from the broad-band spectral energy distribution as sown in
  Fig.~\ref{f:sed} which requires a remaining star of almost zero temperature.
\label{f:tdist}}
\end{figure}

Finally the modeled light curve is normalized to the observed 
curve by a 
$(R_{\infty}/D)^2$ factor. For our model 'A' we derive a distance $D$
of only 76\,pc for an assumed $R_{\rm ns} = 12$\,km star ($R_\infty = 13.8$\,km
for $r_{\rm gr} = 4$). Our best-fit model curve 
for model 'A' is shown in
Fig.~\ref{f:2sm}. The corresponding temperature variation along stellar
latitude is shown in Fig.~\ref{f:tdist} for an assumed atmospheric temperature
of 42 eV.
{It is emphasized again that the fit of the light curve is based
  on an adaption per eye and not due to a formal least-squares fit.}

Having set the main scenery with our {\it ad hoc} two-spot model, we proceed
by comparing the observed light curves with the predictions for the
temperature distribution in a magnetized neutron star crust. Such
distributions became recently available (Geppert et al.~2004) and these
authors kindly provided tabulated data. Geppert et al.~calculated the 
temperature distributions in the atmosphere for the 'core' and
'crustal' field scenario. They solve the equation of heat transport through
the star's atmosphere in the presence of a given dipolar field, the main
parameters which determine the temperature distribution being the (isothermal)
core temperature $T_c$ and the dipolar field strength. The $T$-distribution of
their crustal models for sufficiently high magnetic field strength resembles
the situation studied here with two warm spots. Moreover, the maximum
temperature of their $T_c = 10^8$\,K model is of the same order as the spot
temperature in RBS1223. The 'core' field scenario a la Greenstein \& Hartke
(1983) always produces a too flat $T$-distribution and is not further
considered here. 

\begin{figure}[t]
\resizebox{\hsize}{!}{
\includegraphics[clip=]{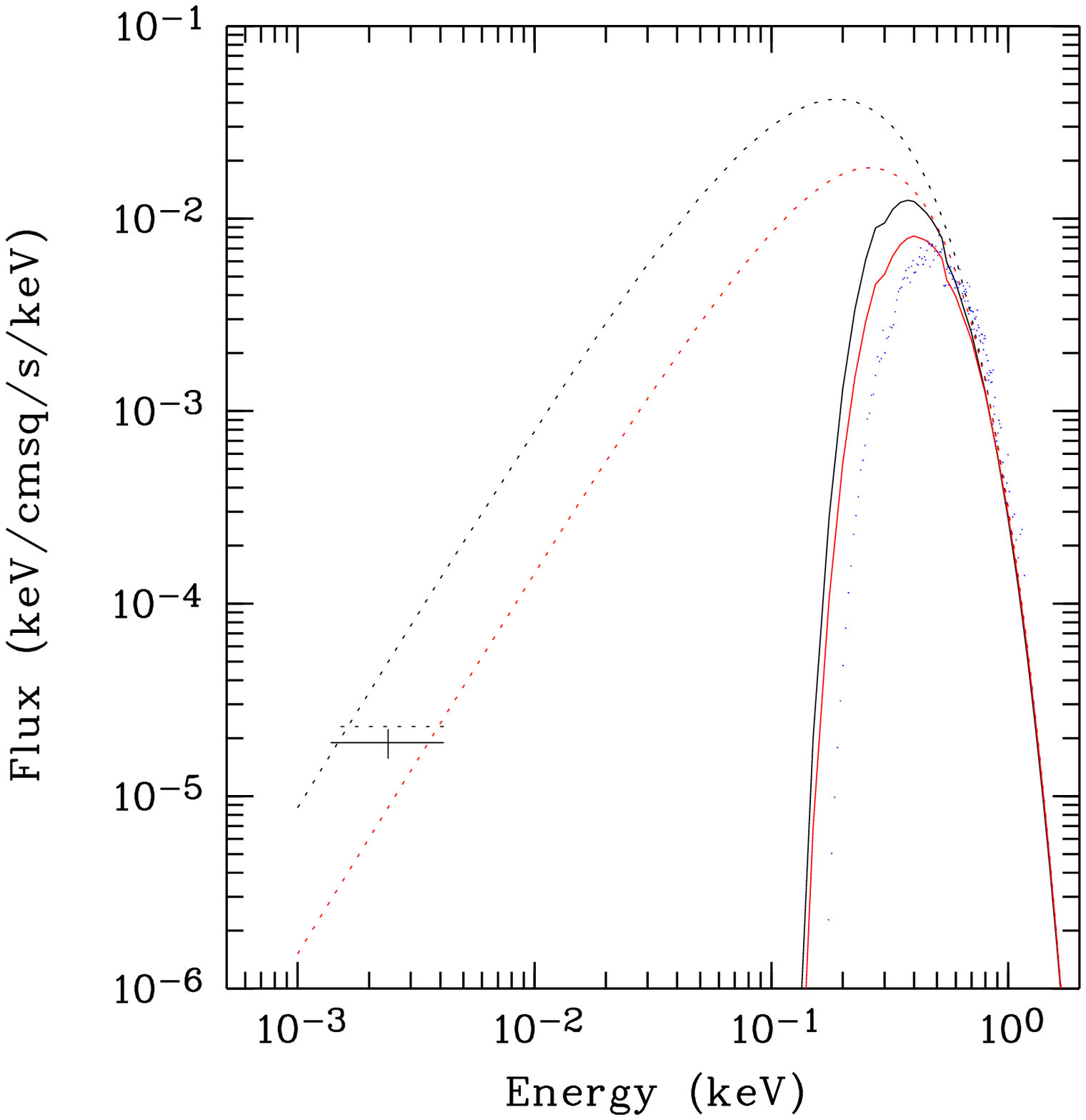}}
\caption{Broad-band spectral energy distribution of RBS1223. 
Shown with solid lines are the absorbed blackbody spectra {which agree}
the XMM-Newton EPIC pn data at phase of the main hump, 
with dotted lines the unabsorbed extrapolations to low
energies for models A (lower red lines) and B (upper black lines),
respectively. The data point at 2.4 eV represents the $m_{\rm 50CCD} = 28.56
\pm0.13$\,mag tentative optical counterpart (Kaplan et al.~2002a). The
unfolded observed X-ray spectrum at phase of the main hump is shown with small
blue symbols. 
\label{f:sed}}
\end{figure}

In order to compute synthetic lightcurves we used the published
$T$-distributions from Geppert et al.~(plus one unpublished distribution for
$T_c = 10^8$\,K and $B = 10^{14}$\,G). 
We parameterized the maximum temperature as a function of given $T_c$ and $B$
using a bi-polynomial interpolation. This allows an easy
scaling of a chosen $T$-profile to the observed maximum temperature.
For light curve synthesis we used the same viewing and magnetic geometry and
the same compactness parameter $r_{\rm  gr}$ 
as above for our initial toy model 'A'. 
The star was divided into two
half-spheres so that different $T$-distributions could
be applied to the two halfs. This somewhat artificial approach was
necessary in order to simulate the two observed unequal spots. 
The two half spheres were slightly inclined with respect to each
other, in order to match the observed phases of the light curve maxima and
minima. 

The normalized $T$-profiles chosen for the main and secondary humps
correspond to $T_c = 10^8$\,K, $\log B \mathrm{ (G)} = 14$, and
$T_c = 10^8$\,K, $\log B \mathrm{ (G)}=13.7$, respectively.

A maximum temperature of 98\,eV in the main spot was achieved  
assuming a core temperature $T_c = 10^{8.04}$\,K 
and a dipolar field of $\log B \mathrm{ (G)} = 13.0$.
In order to model the observed lower
temperature in the second spot, we had to assume a slightly lower temperature
$T_c = 10^{8.02}$\,K and a slightly lower field strength $\log B = 12.5$\,G as
well. This parameter combination then yields the maximum temperature in the
spot of 91\,eV. 
The fit based on the described combination of parameters 
is shown in Fig.~\ref{f:cfm}. As above, we used a radius of the neutron star
$R_\infty = 13.8$\,km. The implied distance derived for the crustal field
model is then 380\,pc. This value is clearly much larger than that of the
simple two-spot model since a much larger fraction of the star's surface
contributes to the observed light. 

Hence, the crustal field temperature profiles seem to be equally suited 
to reflect the observed light curves of RBS1223. We note, however, that our approach
is not selfconsistent, since the maximum spot temperature and the $T$-profiles were
adapted in separate steps. In addition, the simulated $T$-profiles
were computed for a more compact star, $R_{\rm ns} = 10$\,km, $M =
1.4$\,M$_\odot$, than assumed here. 
This might explain the inconsistency
between the values of the parameters $T_c,B$ used to determine the maximum
temperature in the spots and the corresponding $T$-profile used by us.
The temperature
distribution which matches the observed narrow pulses is steeper
than implied by the maximum temperature in the spot. 


\section{Results and discussion}
We have presented a multi-mission timing analysis of the isolated neutron star
RBS1223. New observations with XMM-Newton gave further support for a revised
spin period at 10.31\,s. The spin history could not yet be uncovered without
ambiguity, the previously derived significant spin-down however seems to be
ruled out by the series of XMM-Newton observations. This finding most probably
rules out a magnetic field strength in excess of $10^{14}$\,G. A field
estimate based on the energy loss of a rotating vacuum dipole magnetosphere
$B\sin{\vartheta} \simeq 10^{12} \sqrt{P\dot{P}_{\rm -13}}$ reveals 
with  $P = 10$\,s,  $\dot{P} = 4 \times 10^{-13}$\,s\,s$^{-1}$ and $\vartheta =
80\degr$ a field strength of $B \simeq 6 \times 10^{13}$\,G.
However, the present
data even do not reveal clearly a spin-down of the star. The omission of the
least significant spin value based on the 1996 ROSAT observation is indicative
even of a spin-up of the star. 
The unequivocal determination of the spin history of
RBS1223 without cycle count alias 
is possible by a dedicated campaign with XMM-Newton which is already
granted.

The long-term observations and the analysis presented here shed some new light
on the likely nature of RBS1223. We mention firstly, that there is no
long-term change of the brightness of RBS1223. This makes Bondi-Hoyle
accretion as powering mechanism of the X-ray source unlikely. 

The spin-phase averaged light curve is
double-humped with two humps of different count rate and spectral
hardness. The humps are separated by about 0.47 phase units, the minima by
about 0.43 phase units, which indicates a slight asymmetry of the shape of the
individual humps. The double-humped light curves is indicative for the
presence of a moderately strong field which is not axisymmetric.
A displaced dipole or contribution(s) from higher multipoles are
likely causes for the asymmetry.

We performed a phase-resolved X-ray spectral analysis of the two calibration
observations in 2003 and 2004. A successful fit to the spectra at all phases
was achieved by the combined black-body plus Gaussian absorption line
model. There is a clear temperature variation over the spin cycle, also the
line centre, the line flux and the line equivalent width show cyclic changes. 
The combined new data do not allow to constrain the nature of the Gaussian
further, we still regard the cyclotron absorption line scenario as possible.

The observed variations of the spectral parameters are, however,
different from those observed in RX\,J0720.4--3125 (henceforth RX0720) by Haberl et
al.~(2004). In RX0720 the temperature variations are much smaller,
$\Delta T \simeq 2.5$\,eV only. The equivalent widths of the putative
cyclotron absorption lines change in both systems by about 50\%, the
equivalent width in RBS1223 is a factor 2--3 larger than in RX0720. 
While RX0720 shows a trend that larger equivalent width corresponds
with lower blackbody temperature there is no such trend in RBS1223. 

The double-humped light curve is suggestive of a spotted atmosphere of the
star with two spots separated by about 160 degrees. We presented two
similarly successful fits to the X-ray light curves in two energy
bands. The chosen energy bands for our experiment,  
0.6--0.8, and 0.8--1.0\,keV  respectively,
are thought to be unaffected by the absorption feature 
so that simple blackbody spectra could be used for light curve
synthesis. Taking the view
of the simple model 'A', the star is described by two visible spots of size
8\degr\ and 10\degr\ and a, for X-ray eyes, invisible remaining surface. 
The atmospheric temperature of less than 42\,eV (500000\,K) leads to a revised
age estimate of about $10^{5\dots6}$\,years (derived from cooling curves of
Lattimer \& Prakash 2004). This revision solves the problem between the
estimated young age and the non-detection of an associated  supernova remnant
(paper 2). 

In Fig.~\ref{f:sed} the broad-band spectral energy distribution of
RBS1223 is shown. The best-fit X-ray spectra for the main hump based
on our models 'A' and 'B' (mean spectra of the visible hemispheres) 
are extrapolated to the optical. 
The data point in the optical wavelength range at 2.4 eV
stretching from 3000 to 9000\,\AA\ represents the brightness of the 
$m_{\rm 50CCD} = 28.56
\pm0.13$\,mag tentative optical counterpart dicovered by Kaplan et
al.~(2002a). The extrapolation indicates that the brightness of the likely
optical counterpart is compatible with our models. Any further low
temperature component would raise the predicted flux at optical
wavelength over the observed value. However, the blackbody models which were
adjusted in the energy bands 0.6-0.8, and 0.8-1.0 keV, respectively, already
overpredict the X-ray flux below 0.5\,keV were the Gaussian absorption line is
observed. The extrapolation into the optical may therefore be done only with
caution.

The predicted distance to RBS1223 based on model 'A' is
uncomfortably short, 76\,pc, i.e.~nearer than the much brighter prototypical
system RX\, J1856.5-3754 ($D_{\rm RXJ1856} = 140 \pm 40$\,pc, Kaplan
et al.~2002b). The short distance is due to the simplicity of the
temperature structure and geometry. Any temperature profile with a
more gradual variation of the temperature and corresponding larger
parts of the stellar surface contributing to the observed light will
requires larger distances to the star.
Model 'A' was used as a toy model in order to constrain
the geometry and the compactness, but it lacks a physical interpretation. 

The physical interpretation is possible with our model 'B', 
based on the crustal field model by Geppert
et al.~(2004). This reflects the data equally well. 
In this model the spots are more extended,
i.e.~larger parts of the atmosphere contribute to the observed
radiation. The estimated distance is consequently much larger, $D \sim
380$\,pc (all distance estimates were made for a 1\,M$_\odot$ star with radius
$R_{\rm ns} = 12$\,km and $R_{\rm ns}/r_{\rm S} = 4$). 
However the parameter set based on model
'B' is not completely consistent. 
The average temperature of the
atmosphere is higher than for model 'A', 55\,eV, but still clearly lower than
previously assumed so that the age problem is solved also with model 'B'.

Given the very restricting assumptions of our spectral model, pure
black-body emission with a simple foreshortening law, 
the results of the study presented here can be regarded only as a first
step towards a full model of the star. As worked out by Pavlov et al.~(1996)
and re-addressed by Zavlin \& Pavlov
(2002) atmosphere model fits give temperatures $T_{\rm atm}$ significantly
lower than the blackbody temperature $T_{\rm bb}$ (dependent on the chemical
composition). Light-element models, however, which would show the most drastic
effect are likely ruled out, since they predict much higher fluxes in the
RJ-part of the spectrum. They also would predict much shorter distances
compared to bb-models, which doesn't seem very likely.
Fe-models, on the other hand, do not show dramatic
differences as far as $T_{\rm eff}$ and the predicted distance are
concerned. Since they produce less optical flux they seem to be much better
suited to fit the spectral energy distribution than bb- or light element
models. Another point of uncertainty is the type of the assumed
foreshortening. Zavlin \& Pavlov (2002) computed the angular characteristic of
specific intensities with an assumed field perpendicular to the surface which
shows strongly peaked emission normal to the surface. If applicable to the
case of RBS1223 this would clearly affect the spin-phased light curves.
Future light curve models would
strongly benefit from the availability of a dense grid of models for
different temperature, magnetic field strength, chemical composition and 
angle between the local magnetic field and the surface normal. Based on such
grid, light curve modeling would be possible via a regularisation scheme in
the multidimensional parameter space.

In sum, the revised value for $\dot{P}$ together with the revised value for
the mean atmospheric temperature suggest a nature of RBS1223 as a medium-aged
neutron star ($10^{5\dots6}$\,yrs) on its cooling track. The nature of the
Gaussian is not understood, it could be a cyclotron absorption line in a field
of a few $10^{13}$\,G, still compatible with the (uncertain) spin-down. The
present distant estimates make the star similarly close as the prototype
RXJ1856, which has a large observed proper motion and a measured parallax. 
Although very much fainter, $m_{\rm 50CCD} = 28\fm6$ (Kaplan et al.~2002a), 
a similar measurement
for RBS1223 seems to be feasible and rewarding in order to further constrain
the likely distance, radius and thus nature of this star.

\begin{acknowledgements}
We thank U.R.M.E.~Geppert for providing tabulated temperatures of his
crustal field models and for fruitful discussions.
VVH is supported by the Deutsches Zentrum f\"ur Luft- und Raumfahrt (DLR)
under contract no.~FKZ 50 OX 0201.
\end{acknowledgements}

\end{document}